\documentclass[11pt]{article}
\input{psfig.sty}
\begin{document}
%%%%%%%%%%%%%%%%%%%%%%%%%%%%%%%%%%%%%%%%%%%%%%%%%%%%%%%%%%%%%%%%%%%

\title{
\textbf{Monte Carlo test of critical exponents
in 3D Heisenberg and Ising models}
}

\author{J. Kaupu\v{z}s
\thanks{E--mail: \texttt{kaupuzs@latnet.lv}} \\
Institute of Mathematics and Computer Science, University of Latvia\\
29 Rainja Boulevard, LV--1459 Riga, Latvia}

\date{\today}

\maketitle

\begin{abstract}
We have tested the theoretical values of critical exponents,
predicted for the three--dimensional Heisenberg model,
based on the published Monte Carlo (MC) simulation data for
the susceptibility. Two different sets of the critical exponents
have been considered -- one provided by the usual (perturbative)
renormalization group (RG) theory, and another predicted by
grouping of Feynman diagrams in $\varphi^4$ model (our theory).
The test consists of two steps. First we determine the critical
coupling by fitting the MC data
to the theoretical expression, including both confluent and analytical
corrections to scaling, the values of critical exponents being taken
from theory.
Then we use the obtained value
of critical coupling to test the agreement between
theory and MC data at criticality. As a result, we have found
that predictions of our theory ($\gamma=19/14$, $\eta=1/10$,
$\omega=3/5$) are consistent, whereas those of the perturbative
RG theory ($\gamma \simeq 1.3895$, $\eta \simeq 0.0355$,
$\omega \simeq 0.782$) are inconsistent with the MC data.
The seemable agreement between the RG prediction for $\eta$ and
MC results at criticality, reported in literature, appears due to slightly
overestimated value of the critical coupling.
Estimation of critical exponents of 3D Ising model from complex
zeroth of the partition function is discussed. A refined analysis
yields the best estimate $1/\nu \simeq 1.518$. We conclude that
the recent MC data can be completely explained within our 
theory (providing $1/\nu=1.5$ and $\omega=0.5$)
rather than within the conventional RG theory.
\end{abstract}


\section{Introduction}
In our previous work~\cite{my3}, we have reported the possible
values of exact critical exponents for the Ginzburg--Landau phase
transition model predicted from a reorganized perturbation theory.
These predictions are in exact agreement with the known exact and
rigorous results in two dimensions~\cite{Baxter}, and are equally
valid also in three dimensions.
Our predictions have been compared to some original data of
Monte Carlo (MC) simulations and experiments~\cite{IS,SM,GA},
and a remarkable agreement has been found.

However, there is still rather paradoxical and unclear situation
regarding the MC results. On the one hand, we have
shown theoretically~\cite{my3} the invalidity of the
conventional RG expansions~\cite{Wilson,Ma,Justin},
but, on the other hand, the published
papers on MC simulations usually claim to confirm the values of critical
exponents coming from these expansions and being in contradiction
to our results.

Contrary to the usual claims in the published papers
that the values of critical exponents can be obtained from the
Monte Carlo data with a striking accuracy, i.~e. with an
error much smaller than $0.01$, our expierence in analysis
of several such data shows that in reality it is very difficult
to extract accurate and reliable estimates therefrom.
The problem is that a fitting of MC data to a simple theoretical
ansatz (including no corrections to scaling) can provide a rather
small statistical error, but the obtained result is not reliable
since it contains an uncontrolled systematical error due to
the neglected corrections to scaling. Moreover, confluent
(i.~e., those related to the universal properties of the critical
point) and analytical corrections can be equally
important at finite values of the reduced temperature
at which the simulations have been done, since the amplitude
of the leading analytical correction can be remarkably larger than
that of the confluent correction. Our analysis of the
susceptibility data for the three--dimensional Heisenberg model
(Sec.~\ref{sec:gam}) has shown that
the estimated value of the critical exponent $\gamma$ decreases
by several percents due to the confluent correction, and the
result can be changed remarkably by the analytical correction
too. Thus, both kind of corrections should be taken
into account, but this is not possible in the usual
applications related to the determination of
critical exponents, since inclusion of both kind of corrections in a
theoretical ansatz strongly increases the statistical errors.

 As regards the fitting of MC data at
criticality, only confluent corrections are present, but
the usual estimations are rather sensitive to the precise
value of the critical coupling.
In this aspect, our estimation of the critical exponent
$\eta$~\cite{my3} from the MC simulated fractal
dimensionality of the three--dimensional Ising model
at the critical point
(i.~e., from MC data of~\cite{IS}) is preferable
to a more conventional, but much more sensitive
to the precise value of the critical coupling $\beta_c$, estimation
of this exponent from the susceptibility data at criticality.
 According to the published results~\cite{Janke},
the second method seems to give smaller values of $\eta$
in three dimensions (about $0.027$ for Heisenberg
model~\cite{Janke}) as compared to the
first one (about $1/8$ for Ising model~\cite{my3}),
but the reason for the discrepancy could be an inaccuracy in the
estimated value of $\beta_c$. In the case of the Heisenberg
model, this value has been overestimated, indeed,
as discussed in Sec.~\ref{sec:coupl}.
More recent MC results reported in~\cite{Ballesteros}
also provide rather small values of $\eta$ (about $0.04$
for $O(n)$ models with $n=2,3,4$).
However, the infinite volume
extrapolation in~\cite{Ballesteros} is erroneous in view of
our theory, and a selfconsistent treatment, based on our theoretical
predictions, reveals no contradiction to the MC data
(Sec.~\ref{sec:crex}).

 In the present work we have proposed a Monte Carlo test,
based on a high quality susceptibility data~\cite{Janke}, where
the above discussed problems with corrections to scaling
are solved on a higher level than
it has been done in the currently published papers.
Namely, our method enables us to test
the agreement of MC data with given (fixed) theoretical values
of critical exponents by taking into account both the leading
confluent and the leading analytical correction.
 Our test consists of a very accurate determination of
the critical coupling followed by a fitting of the susceptibility data
at criticality. It has shown (Sec.~\ref{sec:test}) that the actually
discussed MC data are in agreement with our theoretical values
of critical exponents, but not with those of the RG expansions.

\section{Critical exponents from our theory} \label{sec:crex}

 Our theory provides possible values of exact critical exponents
$\gamma$ and $\nu$ for the $\varphi^4$ model whith $O(n)$
symmetry ($n$--component vector model) with the Hamiltonian
\begin{equation} \label{eq:H}
H/T= \int \left[ r_0 \varphi^2({\bf x})
+ c (\nabla \varphi({\bf x}))^2
+ u \varphi^4({\bf x}) \right] d{\bf x} \; ,
\end{equation}
where $r_0$ is the only parameter depending on temperature $T$,
and the dependence is linear.  At the spatial
dimensionality $d=2, 3$ and $n=1, 2, 3, ...$ these values are~\cite{my3}
\begin{eqnarray}
\gamma &=& \frac{d+2j+4m}{d(1+m+j)-2j} \label{eq:gamma} \; , \\
\nu &=& \frac{2(1+m)+j}{d(1+m+j)-2j} \label{eq:nu} \; ,
\end{eqnarray}
where $m \ge 1$ and $j \ge -m$ are integers. At $n=1$ we have
$m=3$ and $j=0$ to fit the known exact results for the
two--dimensional Ising model. As proposed in Ref.~\cite{my3},
in the case of $n=2$ we have $m=3$ and $j=1$, which yields in
three dimensions $\nu=9/13$ and $\gamma=17/13$.

In the present analysis the correction--to--scaling
exponent $\theta$ for the susceptibility also is relevant. The susceptibility
is related to the correlation function in the Fourier representation
$G({\bf k})$, i.~e., $\chi \propto G({\bf 0})$~\cite{Ma}. In the
thermodynamic limit, this relation makes sense at $T > T_c$, where
$T_c$ is the critical temperature.
According to our theory, $G({\bf 0})$ can be expanded in a Taylor
series of $t^{2 \nu -\gamma}$ at $t \to 0$.
In this case the reduced temperature $t$ is defined as
$t=r_0(T)-r_0(T_c) \propto T-T_c$.
Formally, $t^{2 \gamma - d \nu}$ appears as second expansion
parameter in the derivations in Ref.~\cite{my3}, but,
according to the final result represented by
Eqs.~(\ref{eq:gamma}) and~(\ref{eq:nu}),
$(2 \gamma - d \nu)/(2 \nu -\gamma)$ is a natural number.
Some of the expansion coefficients can be zero, so that in general we have
\begin{equation} \label{eq:Delta}
\theta=\ell \, (2 \nu -\gamma) \; ,
\end{equation}
where $\ell$ may have integer values 1, 2, 3, etc. One can expect
that $\ell=4$ holds at $n=1$ (which yields $\theta=1$ at $d=2$ and
$\theta=1/3$ at $d=3$) and the only nonvanishing
corrections are those of the order $t^{\theta}$, $t^{2 \theta}$,
$t^{3 \theta}$, since the known corrections to scaling for
physical quantities, such as magnetization or correlation length,
are analytical in the case of the two--dimensional Ising model.
Here we suppose that the confluent corrections become analytical,
i.~e. $\theta$ takes the value $1$, at $d=2$.
Besides, similar corrections to scaling are expected for
susceptibility $\chi$ and magnetization $M$ since both these
quantities are related to $G({\bf 0})$, i.~e.,
$\chi \propto G({\bf 0})$ and $M^2=\lim_{x \to \infty}
\langle \varphi({\bf 0}) \varphi({\bf x}) \rangle
= \lim_{V \to \infty} G({\bf 0})/V$
hold where $V=L^d$ is the volume and $L$ is the linear size of
the system. The above limit is meaningful at $L \to \infty$,
but $G({\bf 0})/V$ may be considered as a definition of $M^2$
for finite systems too. The latter means that corrections
to finite--size scaling for $\chi$ and $M$ are similar at $T=T_c$.
According to the scaling hypothesis and finite--size scaling
theory (Sec.~\ref{sec:gam}),
the same is true for the discussed here corrections at $t \to 0$.
Thus, the expected expansion of the susceptibility $\chi$ looks
like $\chi = t^{-\gamma} \left( a_0+a_1 t^{\theta} +a_2 t^{2 \theta}
+ \cdots \right)$.

Our general hypothesis is that $j=j(n)$ and $\ell=\ell(n)$
monotoneously increase with $n$ to fit the known exponents
for the spherical model at $n \to \infty$.
In particular, we expect that $j(n)=n-1$,
$\ell(n)=n+3$, and $m=3$ hold at $n=1, 2, 3$ and, probably,
also in general. This hypothesis is well confirmed by MC results
discussed here and in Ref.~\cite{my3}.

We allow that different $\ell$  values correspond to
the leading correction--to--scaling exponent for different
quantities related to $G({\bf k})$. The expansion of
$G({\bf k})$ by itself contains a nonvanishing term of order
$t^{2 \nu -\gamma} \equiv t^{\eta \nu}$ (in the form
$G({\bf k}) \simeq\linebreak t^{-\gamma} \left[ g({\bf k} t^{-\nu})
+ t^{\eta \nu} g_1({\bf k} t^{-\nu}) \right]$ whith
$g_1({\bf 0})=0$, since $\ell >1$ holds in the case of susceptibility)
to compensate the corresponding correction term (produced
by $c \left( \nabla \varphi \right)^2$) in the equation
for $1/G({\bf k})$ (cf.~\cite{my3}).
The latter means, e.~g., that the correlation
length $\xi$ estimated from an approximate ansatz like
$G({\bf k}) \propto 1/ \left[{\bf k}^2+ (1/\xi)^2 \right]$
used in~\cite{Janke,Ballesteros} also contains a correction
proportional to $t^{\eta \nu}$. Since $\eta \nu$ has a rather small value,
the presence of such a correction (and, presumably, also the higher order
corrections $t^{2 \eta \nu}$, $t^{3 \eta \nu}$, etc.) makes the above
ansatz unsuitable for an accurate numerical correction--to--scaling
analysis. Due to this reason the susceptibility data,
but not the correlation length data of Ref.~\cite{Janke},
are used in our further analysis.

The correction $t^{\eta \nu}$ is related to the correction $L^{-\eta}$ 
in the finite--size scaling expressions at criticality.
 The existence of such a correction in the asymptotic
expansion of the critical real--space Green's (correlation) function,
i.~e.\linebreak $\widetilde G(rL) \propto L^{2-\eta-d} \left(1 + o(L^{-\eta}) \right)$
where $r$ is a constant, is well confirmed by our recent (preliminary)
results for the 2D~Ising model. These results for
$L=2, 4, 6, \ldots, 16$ have been obtained by an exact numerical
transfer--matrix algorithm.
  In such a way, the infinite volume extrapolation in~\cite{Ballesteros}
appears to be incorrect, therefore the obtained there
results do not represent a serious argument against our theory.
Moreover, if the extrapolation in~\cite{Ballesteros} is done
including the correction $L^{-\eta}$, then the results for
$O(n)$ models with $n=2, 3$ appear to be in a satisfactory agreement
(within the extrapolation errors) 
with our values $\eta=1/9$ and $\eta=1/10$, respectively.

  Our consideration can be generalized easily to the case
where the Hamiltonian parameter $r_0$ is a nonlinear analytical
function of $T$. Nothing is changed in the above expansions
if the reduced temperature $t$, as before, is defined by
$t=r_0(T)-r_0(T_c)$. However, analytical corrections to scaling appear
(and also corrections like\linebreak $(T-T_c)^{m+n \theta}$ with integer $m$ and
$n$) if $t$ is reexpanded in terms of $T-T_c$ at $T>T_c$. The
solution at the critical point remains unchanged, since the phase
transition occurs at the same (critical) value of $r_0$.

\section{Estimation of the critical exponent
$\gamma$ from MC data} \label{sec:gam}

 In this section we discuss the estimation of the susceptibility
exponent $\gamma$ for the classical three--dimensional Heisenberg
model. Our analysis is based on the fitting of
the susceptibility (MC) data to a theoretical ansatz.
 According to the finite--size scaling theory, the susceptibility
$\chi$ depending on the reduced temperature
\linebreak $t= 1 - \beta / \beta_c$ (where $t>0$) and the linear size of
the system $L$ reads
\begin{equation} \label{eq:sc1}
\chi = L^{\gamma/\nu} g \left( L/\xi \right) \; ,
\end{equation}
where $g(L/\xi)$ is the scaling function and $\xi \sim t^{-\nu}$
is the correlation length of an infinite system. Eq.~(\ref{eq:sc1})
holds precisely at $L \to \infty$ and $t \to 0$ for any given
value of $L/\xi$. At finite values of $t$ and $L$ corrections
to~(\ref{eq:sc1}) exist. Eq.~(\ref{eq:sc1}) can be rewritten as
\begin{equation} \label{eq:sc2}
\chi = t^{-\gamma} f \left( t L^{1/\nu} \right) \; ,
\end{equation}
where $g(y)=y^{-\gamma/\nu} f \left( y^{1/\nu} \right)$.
In the thermodynamic limit $L \to \infty$ Eq.~(\ref{eq:sc2})
reduces to $\chi = b_0 \, t^{-\gamma}$,
where $b_0 = \lim\limits_{x \to \infty} f(x)$
is the amplitude. A natural extension of Eq.~(\ref{eq:sc2}),
including corrections to scaling, is
\begin{equation} \label{eq:exp}
\chi = t^{-\gamma} \sum\limits_{l \ge 0} t^{\gamma_l}
f_l \left( t L^{1/\nu} \right) \; ,
\end{equation}
where $\gamma_0 \equiv 0$, $f_0(x) \equiv f(x)$,
 and the terms with $l>0$ represent
all the corrections in the asymptotic expansion of $\chi$ at $t \to 0$
for any given value of $x = t L^{1/\nu}$. In the thermodynamic
limit we have $\lim\limits_{x \to \infty} f_l(x) =b_l$, where
$b_l$ are the amplitudes. The most important correction terms
in the sum over $l$ are the leading confluent correction $b_1 t^{\gamma_1}$
with the exponent $\gamma_1=\theta$ and the leading analytical
correction $b_2 t^{\gamma_2}$ with $\gamma_2=1$.
Although $\theta<1$ holds, the analytical correction also
should be included at finite values of $t$ used in practical
simulations: because of absence of a direct correlation between
the amplitudes of confluent and analytical corrections, the
ratio $r=b_2/b_1$ can be arbitrarily large.
One can expect that the higher order confluent corrections
(i.~e., those proportional to $t^{2 \theta}$,
$t^{3 \theta}$, etc.) are small as compared to
the leading confluent correction, and the same is true
for analytical corrections.
We consider the case of small $t$ and large $x$, i.~e., small $f_l(x)-b_l$.
In this case Eq.~(\ref{eq:exp}) can be written as
\begin{equation} \label{eq:sc3}
\chi \simeq t^{-\gamma} \left[ 1 + b \,
\left( t^{\theta} +r t \right) \right] \,
f \left( t L^{1/\nu} \right) \; ,
\end{equation}
where $b=b_1/b_0$ is a constant.

 We have used the susceptibility data simulated by an improved
(cluster) MC algorithm reported in Ref.~\cite{Janke} ($\bar \chi^{imp}$
vs $\beta$, tab.~IV in~\cite{Janke}) to estimate the critical
exponent $\gamma$ by fitting the data to~(\ref{eq:sc3}). Such an
estimation has been done in~\cite{Janke},
neglecting either the analytical or the confluent correction
and setting $f \left( t L^{1/\nu} \right)=b_0$.
Since in the actual simulations the scaling argument
$x=t L^{1/\nu}$ has large enough values, about $6$ or $7$, which
are varied only slightly, the latter approximation is
reasonable. We have used even better approximation where $\ln f(x)$
has been linearized within the narrow range of $x$ variation,
and the simulated data points for $\ln \chi$ have been fitted to
the resulting theoretical expression
\begin{equation} \label{eq:fit}
\ln \chi (t,L) = a -\gamma \ln t + \ln \left[ 1+
b \left( t^{\theta} +rt \right) \right] + p \, tL^{1/\nu} \; ,
\end{equation}
where $a$ and $p$ are constants. The minimum of the sum of the
squared deviations for $N$ data points $S(N)$ corresponds to the
least--squares fit. Besides, it is reasonable to use the
least--squares method just for $\ln \chi$, but not for $\chi$,
since the errors for $\ln \chi$ data points are comparable, i.~e.,
the relative but not the absolute errors are more or less equal.
At large $N$, the inaccuracy in the fitted curve due to the statistical
errors can be characterised by the standard deviation
$\sigma = (S(N)/[N(N-1)])^{1/2}$. Obviously, the minimum of $\sigma$
corresponds to the least--squares fit at any given $N$.

\begin{figure}
\centerline{\psfig{figure=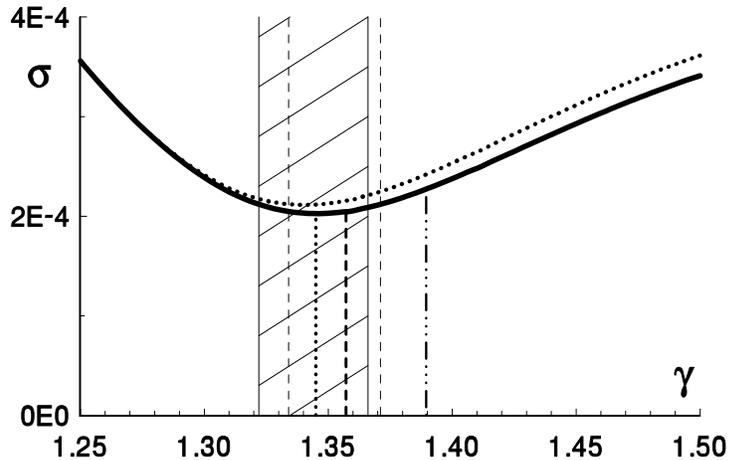,width=11cm,height=8.5cm}}
\vspace{-3ex}
\caption{\small Estimation of the critical exponent $\gamma$ in
 3D Heisenberg model. Solid line shows the standard deviation
$\sigma$ of the simulated data
points from the analytical curve~(\ref{eq:fit})
as a function of $\gamma$ with parameters $a$, $b$,
$p$, and $\beta_c$ obtained from the least--squares fit at 
$\nu=5/7$, $\theta=3/7$ (our theoretical values), and $r=0$.
The dotted curve corresponds to fixed $p=0$. The minimum of the
solid curve gives the least--squres estimate
%of the susceptibility exponent
$\gamma=1.345$. All fits (for different data sets) lie in the marked
area which is shifted only slightly, as indicated by thin vertical dashed
lines, if the RG values of $\nu$ and $\theta$ are used instead of ours.
Our theoretical value $\gamma=19/14$ (thick vertical dashed line)
is inside of the marked region, whereas that of the RG theory
(vertical dot--dot--dashed line at $\gamma=1.3895$) is outside.}
\label{gamfit}
\end{figure}

 We have illustrated in Fig.~\ref{gamfit} the estimation
of $\gamma$ by minimizing $\sigma$ with respect to the parameters
$a$, $b$, $p$, and $\beta_c$
(where $\beta_c$ is incorporated in~(\ref{eq:fit})
via $t= 1- \beta/\beta_c$)
at fixed exponents $\theta=3/7$ and $\nu=5/7$,
taken from our theory (Sec.~\ref{sec:crex}).
The analytical correction to scaling has been neglected by
setting $r=0$. The solid line shows the accuracy of the fit,
i.~e. the value of $\sigma$, depending on the choice
of the exponent $\gamma$. The minimum of $\sigma$, indicated by
a vertical dotted line, is located at
$\gamma \simeq 1.345$, which corresponds to the least--squares fit.
The dotted curve corresponds to the case of fixed $p=0$.
From this we can see that inclusion of the term $p \, tL^{1/\nu}$
in~(\ref{eq:fit}), responsible for the variation of the scaling
function $f(x)$, affects the result only slightly.

In spite of the very high accuracy of the fit (about $0.02 \%$
error in $\chi$), the minimum in $\sigma$ is too broad for a
reliable estimation of $\gamma$ with, e.~g., $\pm 0.01$ accuracy.
This is a problem which usually appears if we use a high--level
approximation including many fitting parameters. If the
analytical correction also is included, then the situation
becomes even worse, i.~e., the $\sigma$ vs $\gamma$ plot
is an almost horizontal line. Neglection of both (confluent
and analytical) corrections, as it has been done finally
in~\cite{Janke}, is not a solution of the
problem since the result is affected significantly by the confluent
correction. Namely, the obtained value of $\gamma$ is shifted from
$1.389$ to $1.345$. According to our estimation, the statistical error
for the latter result is remarkably smaller than the difference between
these two values, so that the second value is better. Another problem is
that the estimated value of
$\gamma$ depends on $\theta$ and $\nu$. This effect, however, is relatively
small. By the conventional RG values $\theta=0.55$ and
$\nu=0.7073$~\cite{Justin1} we obtain $\gamma \simeq 1.354$.

 Like in Ref.~\cite{my3}, we have estimated the possible statistical
error of our result $\gamma \simeq 1.345$ by comparing the values
of $\gamma$ for a large number of different data sets generated
from the original one (with 18 data points) by omitting some
(1 to 6) data points. The data points have been omitted more or
less randomly, but not the neighbouring points and not the first
and the last point simultaneously, to ensure a sufficiently
uniform distribution of the used $t$ values and to avoid a significant
narrowing of the total interval covered by these values. The largest
deviations from the central $\gamma$ value $1.345$ have been observed
omitting the data points No.~1, 6, 10, 14, and 17 (tab.~IV in~\cite{Janke}),
which yielded $\gamma \simeq 1.322$, and the data points No.~2, 5, 8, 11,
and 14, which yielded $\gamma \simeq 1.366$. Thus, all the fits gave
$1.322 \le \gamma \le 1.366$ at $r=0$, $\theta=3/7$ and $\nu=5/7$.
This interval is marked in Fig.~\ref{gamfit} by thin solid lines. At
$\theta=0.55$ and $\nu=0.7073$ the borders of this region are shifted
slightly, as indicated by thin vertical dashed lines.
These manipulations enable us to estimate the possible
statistical error in both cases, i.~e., $\gamma=1.345 \pm 0.023$
at $\theta=3/7$ and $\gamma=1.354 \pm 0.020$ at $\theta=0.55$. These,
in fact, are maximal errors, i.~e., since we never have observed larger
deviations, the probability that the value extracted from exact data
would be outside of the error bars is vanishingly small.

It is a remarkable fact that our theoretical value
$\gamma=19/14 \simeq 1.35714$ (thick vertical dashed line)
lies inside the region of maximal statistical errors, whereas that of the
RG theory, i.~e. $\gamma \simeq 1.3895$ indicated by a do--dot--dashed line,
is clearly outside of this region. This result can be changed
by the analytical correction. However, if the ratio of amplitudes
$r$ in Eq.~(\ref{eq:fit}) is positive, then the least--squares fit
with respect to the
parameters $a$, $b$, $p$, and $\beta_c$ always yields the central
value of $\gamma$ (with all 18 data points included) in the range from
$1.345$ to $1.369$ at $\theta=3/7$ and $\nu=5/7$. Here
$\gamma=1.369 \pm 0.013$ corresponds to the case of purely analytical
correction to scaling obtained by formally setting $\theta=1$.
In such a way, selfconsistent estimations at $r>0$ yield $\gamma$ values
which are reasonably close to our prediction $\gamma=19/14$. 
Precise agreement is reached at $r \simeq 1.17$.

 Unfortunately, we have no proof that $r$ is positive. If we
allow that $r<0$, then a large uncertainty appears. In this case
$\gamma$ can take the values as small as, e.~g., $1.1$ (at $\theta=3/7$,
$\nu=5/7$, and $r \approx -1.35$) and as large as $1.4$
(at $\theta=0.55$, $\nu=0.7073$, and $r \approx -1.8$).
The actual MC data do not allow to find the true value of $r$
(unless we assume that our exponents are true and,
therefore, $r \approx 1.17$), since the standard deviation of the
least--squares fit is almost independent on $r$.

\section{Estimation of the critical coupling}
\label{sec:coupl}

\begin{figure}
\centerline{\psfig{figure=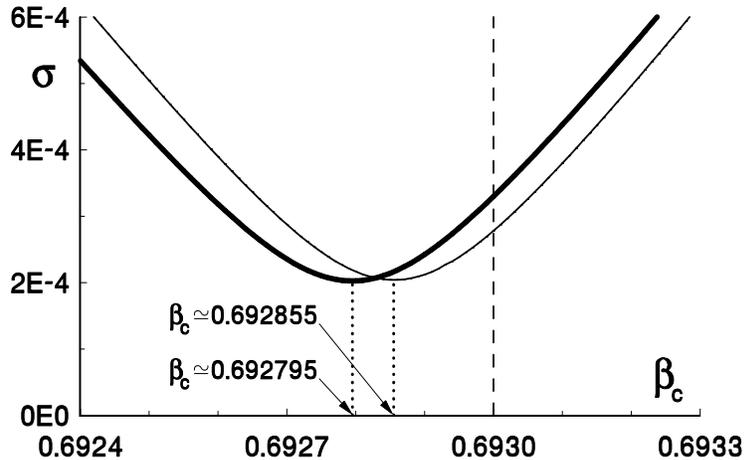,width=11cm,height=8.5cm}}
\vspace{-3ex}
\caption{\small Estimation of the critical coupling $\beta_c$ by fitting
the MC data to ansatz~(\ref{eq:fit}) with fixed exponents taken from
our (thick solid line) and RG (thin solid line) theory. The minimums
of $\sigma$ vs $\beta_c$, used as a fitting
parameter, give the least--squares estimates for the true $\beta_c$ value,
as indicated by vertical dotted lines and arrows. The vertical
dashed line indicates a value of $\beta_c$ proposed
in~\cite{Janke,Ballesteros}.}
\label{betac}
\end{figure}

\begin{figure}
\centerline{\psfig{figure=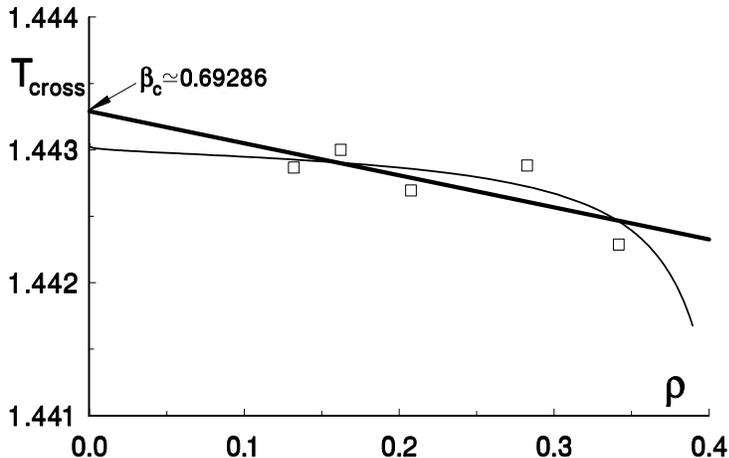,width=11cm,height=8.5cm}}
\vspace{-3ex}
\caption{\small Estimation of the critical coupling by the Binder's cumulant
crossing technique. The straight line represents the
least--squares fit of~(\ref{eq:binder}) to
the MC data for crossing points $T_{cross}$. The zero intercept
gives $\beta_c=1/T_c \simeq 0.69286$. For comparison, the
approximation $T_{cross}-T_c \propto 1/\ln b$ (where $b$ is Binder parameter)
is shown by thin solid line.}
\label{tc}
\end{figure}

 Based on the method developed in Sec.~\ref{sec:gam}, here we determine
the critical coupling $\beta_c$ for the three--dimensional
Heisenberg model assuming that critical exponents
$\gamma$, $\theta$, and $\nu$ are known from theory. The latter ensures
a very small statistical error. The coefficients
in~(\ref{eq:fit}) are found by the least--squares method. The resulting
value of the standard deviation $\sigma$ vs $\beta_c$,
used as a fitting parameter, is shown in Fig.~\ref{betac}.
The thick
solid line corresponds to our critical exponents $\gamma=19/14$,
$\nu=5/7$, and $\theta=3/7$, whereas the thin solid line -- to
the conventional (RG) exponents $\gamma=1.3895$, $\nu=0.7073$, and
$\theta=0.55$. The minimums of these curves, indicated by vertical
dotted lines, are located at $\beta_c \simeq 0.692795$ and
$\beta_c \simeq 0.692855$, respectively, corresponding to the
least--squares estimates for the true values of the critical coupling.
The estimation of maximal statistical errors, as in Sec.~\ref{sec:gam},
leads to the following conclusions:
\begin{enumerate}
\item If our values of the critical exponents $\gamma=19/14$,
$\nu=5/7$, and $\theta=3/7$ are true, then
\begin{equation} \label{eq:betac1}
\beta_c=0.692795 {\mbox{\footnotesize +0.000030}
\atop \mbox{ \footnotesize --0.000043}} \; .
\end{equation}
\item If the true values of the critical exponents are close to
those predicted by the RG theory, i.~e., $\gamma=1.3895$,
$\nu=0.7073$, and $\theta=0.55$, then
\begin{equation} \label{eq:betac2}
\beta_c=0.692855 {\mbox{\footnotesize +0.000029}
\atop \mbox{\footnotesize --0.000043}} \; .
\end{equation}
\end{enumerate}
The estimation in~\cite{Janke,Ballesteros} gave $\beta_c \simeq 0.6930$.
This value is indicated in Fig.~\ref{betac} by a vertical dashed line.
As we see, it clearly does not correspond to the best fit. To clear up
the reason for the discrepancy, let us discuss the Binder's cumulant
crossing technique used in~\cite{Janke} and~\cite{Ballesteros}
for the estimation of $\beta_c$.
In this approach, the magnetization cumulants for different lattice
sizes $L$ and $L'$ are plotted as a function of $\beta$ to find the
intersection point $\beta=\beta_{cross}$. According to the
theory~\cite{Binder},
\begin{equation} \label{eq:binder}
\beta_{cross}(L,b)-\beta_c \propto L^{-(1/\nu)-\omega}
\, \frac{1-b^{-\omega}}{b^{1/\nu}-1}
\end{equation}
holds at large $L$, where $L$ is the size of the smaller lattice,
$b=L'/L$ is the Binder parameter, and $\omega= \theta/\nu$.
The estimation in~\cite{Janke} has been done by approximating
the term $\rho =(1-b^{-\omega})/(b^{1/\nu}-1)$ in Eq.~(\ref{eq:binder})
with $const/\ln b$. We have made and have illustrated in
Fig.~\ref{tc} our own estimation of $\beta_c$ from
Eq.~(\ref{eq:binder}), using the data for $T_{cross}=1/\beta_{cross}$ 
extracted from Fig.~3 in Ref.~\cite{Janke}. 
According to~(\ref{eq:binder}), $\beta_{cross}$ is a linear function of
$\rho$ at a fixed $L$. The same is true for $T_{cross}$
in vicinity of $T_c$. The straight line in Fig.~\ref{tc}
corresponds to the linear least--squares fit for $T_{cross}$ vs
$\rho$ at $L=16$ (with our exponents $\omega=3/5$ and $\nu=5/7$)
which yields (at $\rho=0$) $\beta_c \simeq 0.69286$.
This value agree with~(\ref{eq:betac1}) and~(\ref{eq:betac2})
within the error bars $\pm 0.0001$ proposed in~\cite{Janke}.
In fact, the data points are too much scattered to consider
such an estimation reliable. Due to this reason, we have not
tried to estimate $\beta_c$ from the data of $L=12$ which are
even more scattered. We have depicted in Fig.~\ref{tc}
by thin solid line the fit, corresponding to the approximation
$T_{cross}-T_c \propto 1/\ln b$, made in~\cite{Janke} at $L=16$.
As we see, in the scale where the original ansatz~(\ref{eq:binder})
yields a straight line this approximation is represented by
a curve providing an underestimated value of $T_{cross}$ at $\rho=0$,
i.~e., an overestimated $\beta_c \simeq 0.6930$ instead of
$\beta_c \simeq 0.69286$.
Obviously, this approximation is the reason for the discrepancy. Note
that other kind of estimations in~\cite{Janke} provided
a bit smaller $\beta_c$ values, closer to ours.
%Surprisingly, the same
%overestimated value of $\beta_c$, i.~e.
%$0.6930$, has been obtained both in~\cite{Janke} and~\cite{Ballesteros}.
%Unfortunately, the authors of Ref.~\cite{Ballesteros} have not
%provided enough information to find out the reason for such a striking
%agreement.

\section{The test of consistency at $T=T_c$} \label{sec:test}

 Consequently following the conclusions~(\ref{eq:betac1})
and~(\ref{eq:betac2}) made in Sec.~\ref{sec:coupl}, here we
test the agreement between theory and MC data at criticality.

\begin{figure}
%\vspace*{-2ex}
\centerline{\psfig{figure=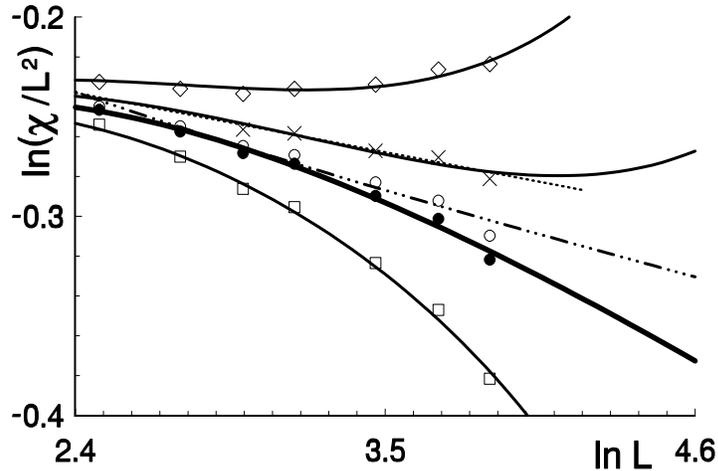,width=11cm,height=8.5cm}}
\vspace*{-3ex}
\caption{\small Our fit to the susceptibility ($\chi$) data of~\cite{Janke}
for 3D Heisenberg model
at and near criticality. Only two coefficients $a$ and $b$
have been used as fitting parameters in~(\ref{eq:etta}) for the
thick solid curve and solid circles representing
$\ln \left( \chi/L^2 \right)$
vs $\ln L$ at criticality according to our critical exponents
($\eta=0.1$, $\gamma=19/14$, $\omega=0.6$) and $\beta_c \simeq 0.692795$
estimated independently in Sec.~\ref{sec:coupl}. The same
fit at RG exponents
%($\eta \simeq 0.0355$, $\gamma \simeq 1.3895$,
%$\theta \simeq 0.55$) 
with the corresponding $\beta_c$ value $0.692855$ is represented by
the dot--dot--dashed line and empty circles. Thin solid lines
show our three--parameter fit at $\beta=0.6925$ (empty squares),
$\beta=0.6930$ (crosses), and $\beta=0.6933$ (empty rhombs). The
linear fit of~\cite{Janke} is shown by tiny dashed line.}
\label{etta}
\end{figure}

According to the finite--size scaling theory, the susceptibility
at the critical point is given by
\begin{equation} \label{eq:crit}
\chi \propto L^{\gamma/\nu} \left( 1+ b L^{-\omega} + \ldots \right) \; ,
\end{equation}
where $b$ and $\omega=\theta/\nu$ are the amplitude and the exponent
of the leading correction to scaling. The dots stand for further
corrections. Eq.~(\ref{eq:crit}) can be rewritten as
\begin{equation}
\label{eq:etta}
\ln \left( \chi /L^2 \right) \simeq a - \eta \ln L
+ \ln \left( 1+ b L^{-\omega} \right) \; ,
\end{equation}
where $a$ is a constant and $\eta=2- \gamma/\nu$ is the critical
exponent describing the asymptotic long--wave behavior of the correlation
function (i.~e. $G({\bf k}) \sim k^{-2+\eta}$) at $T=T_c$.
 We have read from Fig.~6 in Ref.~\cite{Janke} the values of $\chi$
near $\beta_c$ and have made the linear interpolation between
$\beta=0.6927$ and $\beta=0.6929$ to estimate $\chi$ at the
values of the critical coupling given by (\ref{eq:betac1})
and (\ref{eq:betac2}). So obtained $\chi$ values are depicted in
Fig.~\ref{etta} by solid and empty circles, respectively.
The corresponding two parameter ($a$ and $b$ in Eq.~(\ref{eq:etta}))
least--squares fits with fixed exponents
are shown by thick solid line (our case) and dot--dot--dashed line
(RG case). If $\eta$ is considered as a fitting parameter, then in our
case the least--squares fit yields $\eta \simeq 0.105$ in close
agreement with the theoretical value $0.1$, whereas in the RG case
it yields $\eta \simeq 0.076$ in a remarkable disagreement with
the theoretical value $0.0355$. It can be seen also from
Fig.~\ref{etta} that the dot--dot--dashed line
with $a=-0.17034$ and $b=0.1178$, obtained at fixed
$\eta=0.0355$, does not provide a satisfactory fit to the data,
i.e., this line is curved in a wrong direction.

 Our values of critical exponents provide an excellent
fit to the MC data not only at $\beta=\beta_c$, but also at
small deviations $t=1-\beta/\beta_c$ from the critical point
considered in Fig.~6 of Ref.~\cite{Janke}.
Our fit $\chi = 1.1266 \, L^{1.9} \left( 1 -0.4944 L^{-0.6}
-0.58 \, tL^{1.4} \right)$ is shown in Fig.~\ref{etta} by solid
lines. This approximation is consistent with the finite--size
scaling theory at large $L$ and small $tL^{1/\nu}$.
The data points in Fig.~\ref{etta} correspond to $12 \le L \le 48$.
At smaller $L$ values the second--order corrections to scaling,
neglected in our ansatz, could be relevant.
The solid curve at $\beta=0.6930$ is the most linear one
within $12 \le L \le 48$, as it is evident from Fig.~\ref{etta}
where the straight--line fit of~\cite{Janke} is shown by a tiny
dashed line. It is evident also that the good linearity of
$\ln \left( \chi /L^2 \right)$ vs $\ln L$ in this region
does not mean that $\beta_c \simeq 0.6930$ and $\eta \simeq 0.027$.

One of the arguments in~\cite{Janke}, supporting the idea that
$\eta$ has a very small value ($\eta<0.05$), is based
on the simulated data for $\chi$ vs the correlation length
$\xi$ for finite systems. However, the variation
of $\eta_{eff}$ with $\xi$ in Fig.~12 of Ref.~\cite{Janke}
can be well explained by presence of corrections of
the kind $\xi^{-m \eta}$, where $m=1, 2, ...$ and $\eta=1/10$,
consistent with the correction--to--scaling analysis
in Sec.~\ref{sec:crex} (see remarks regarding the actual approximation
for $\xi$). In this case $\eta_{eff}$ can behave nonmonotoneously,
as well.
As regards other arguments in~\cite{Janke} in support of the
conventional RG values of critical exponents, they are weaker
than our contraarguments discussed here, since all the final estimates
in~\cite{Janke} are obtained neglecting corrections to scaling.
 Note also that such kind of simple estimations not always give
very small values of $\eta$. In particular, the values of about
$0.15$ follow from MC study of Heisenberg fluid~\cite{NW}.
In view of our theory and presented here analysis of the MC method,
the discrepancy between the so called "lattice" and "off--lattice"
critical exponents discussed in~\cite{NW} can be well understood
as an error of about $\pm 0.07$ (in $\eta$) of the above discussed
simple estimations.

   It is noteworthy that a large variety of experimental measurements
in Ni discussed in~\cite{St}, confirm our values of critical
exponents $\gamma=19/14=1.357...$,
$\beta=(d-2+\eta)\, \nu/2= 11/28=0.3928...$,
and $\delta=(d+2-\eta)/(d-2+\eta)=49/11=4.4545...$ rather than those
of the RG theory ($\gamma \simeq 1.3895$,
$\delta \simeq 4.794$, and $\beta \simeq 0.3662$~\cite{Justin1}).

\section{Comparison to 3D Ising model}  \label{sec:zeroth}

In this section we discuss the recent MC results~\cite{ADH}
for the complex zeroth of the partition function of the
three--dimensional Ising model. Namely, if the coupling $\beta$
is a complex number, then the statistical sum has zeroth at
certain complex values of $\beta$ or $u= e^{-\beta}$.
The nearest to the real positive axis values $\beta_1^0$ and
$u_1^0$ are of special interest. Neglecting the second--order
corrections, $u_1^0$ behaves like
\begin{equation}
u_1^0 = u_c + A\, L^{-1/\nu} +B\, L^{-(1/\nu)-\omega}
\end{equation}
at large $L$, where $u_c=e^{-\beta_c}$ is the critical value
of $u$, $A$ and $B$ are complex constants, and $\omega$ is the
correction--to--scaling exponent. According to the known
%exact
results (see, e.~g., the solution given in~\cite{Brout}), the
partition function zeroth correspond to complex values of
$\, \sinh (2 \beta) \,$ located on a unit circle
in the case of 2D Ising model, so that $A$ is purely imaginar. The latter
means that the critical behavior of real and imaginary
parts of $u_1^0 - u_c$ essentially differ from each other,
i.~e., $Re \left( u_1^0 -u_c \right) \propto L^{-(1/\nu) -\omega}$
and $Im \left( u_1^0 \right) \propto L^{-1/\nu}$
(where, in this case of $d=2$, $\nu=\omega=1$) at $L \to \infty$.
The MC data of~\cite{ADH}, in fact, provide a good evidence that
the same is true in three dimensions.
 
 Unfortunately, the authors of Ref.~\cite{ADH} have not tried
to find the objective truth regarding the behavior of the
complex zeroth, but only have searched the way how to confirm the
already known estimates for $\nu$. Their treatment, however, is
rather doubtful. First, let us mention that, in contradiction to
the definition in the paper, $u_1^0$ values
listed in Tab.~I of~\cite{ADH} are not equal to $e^{-\beta_1^0}$
(they look like $e^{-4\beta_1^0}$).
%, i.~e., they contradict the definition in the paper.
%They look like $e^{-4\beta_1^0}$, but there is no reason
%to consider $e^{-4\beta_1^0}$ instead of $e^{-\beta_1^0}$,
%unless it helps to obtain some a priori known desired value of $\nu$.
Second, the fit to a theoretical ansatz for $\mid u_1^0(L) - u_c \mid$,
Eq.~(6) in~\cite{ADH}, is unsatisfactory. This ansatz contains a
misterious parameter $a_3$. If we compare Eqs.~(5) and (6)
in~\cite{ADH}, then we see immediately that
$a_3 \equiv (1/\nu)+\omega$. At the same time, the obtained estimate
for $a_3$, i.~e. $a_3 =4.861(84)$, is completely inconsistent with
the values of $(1/\nu)+\omega$, about $2.34$, which follow from
authors own considerations. Our prediction, consistent with
the correction--to--scaling analysis in Sec.~\ref{sec:crex}
(and with $\ell=4$ in~(\ref{sec:crex}) to coincide with the known
exact result at $d=2$), is $\nu=2/3$ and $\omega=1/2$, i.~e.,
$(1/\nu)+\omega=2$.

\begin{figure}
\centerline{\psfig{figure=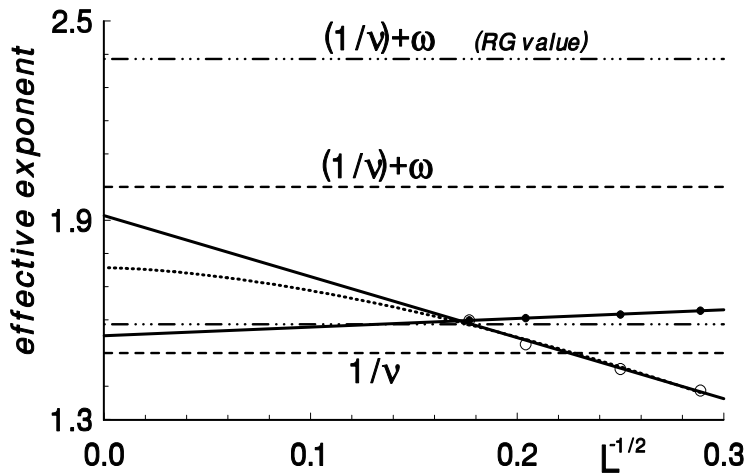,width=11cm,height=8.5cm}}
\vspace{-3ex}
\caption{\small Effective critical exponents for the real (empty circles)
and the imaginary (solid circles) part of complex
partition--function--zeroth of 3D Ising model depending on
$L^{-1/2}$, where $L$ is the linear size of the system.
Solid lines show the linear least--squares fits. 
The asymptotic values from our theory are indicated by horizontal
dashed lines, whereas those of the RG theory -- by
dot--dot--dashed lines. A selfconsistent extrapolation within
the RG theory corresponds to the tiny dashed line.}
\label{zeroth}
\end{figure}

  To obtain a more complete picture, we have considered separately
the real part and the imaginary part of $u_1^0-u_c$. We have calculated
$u_1^0$ from $\beta_1^0$ data listed in Tab.~I of~\cite{ADH} and
have estimated the effective critical exponents $y_{eff}'(L)$
and $y_{eff}''(L)$, separately
for $Re \left( u_1^0-u_c \right)$ and $Im \left( u_1^0 \right)$,
by fitting these quantities to an ansatz $const \cdot L^{-y_{eff}'}$
and $const \cdot L^{-y_{eff}''}$, respectively,
at sizes $L$ and $L/2$. The value of $u_c$ consitent with
high-- and low-- temperature series~\cite{SA}
as well as MC~\cite{FL} estimations
of the critical coupling, $\beta_c \simeq 0.221659$, have been used.
The results are shown in Fig.~\ref{zeroth}.
As we see, $y_{eff}'$ (empty circles) claims to increase
above $y_{eff}''$ (solid circles) when $L$ increases.
This is a good numerical evidence that, like in the two--dimensional
case, the asymptotic values are
$y'=\lim_{L \to \infty} y_{eff}'(L)=(1/\nu)+\omega$ and
$y''=\lim_{L \to \infty} y_{eff}''(L)=1/\nu$.
 According to our theory, the actual plots in the $L^{-1/2}$ scale
are linear at $L \to \infty$, as consistent with the expansion
in terms of $L^{-\omega}$. The linear least--squares fits are shown
by solid lines. The zero intercepts $1.552$ and $1.913$ are in approximate
agreement with our theoretical values $1.5$ and $2$ indicated by horizontal
dashed lines. The relative discrepancy of about $4\%$, presumably, is due
to the extrapolation errors and inaccuracy in the simulated data.

The behavior of $y_{eff}'$ is rather inconsistent with
the RG predictions. On the one hand, $y_{eff}'$ claims to increase
above $y_{eff}''$ and also well above the RG value
of $1/\nu$ (the lower dod--dot--dashed line at $1.5863$), and, on the other
hand, the extrapolation yields $y'$ value ($1.913$) which is remarkably
smaller than
$(1/\nu)+\omega \simeq 2.3853$ (the upper dot--dot--dashed line)
predicted by the RG theory. For selfconsistency,
we should use the linear extrapolation in the scale of $L^{-\omega}$ with
$\omega=0.799$ (the RG value). However, this extrapolation
(tiny dashed line in Fig.~\ref{zeroth}), yielding
$y' \simeq 1.757$, does not solve the problem
in favour of the RG theory.

\begin{figure}
\centerline{\psfig{figure=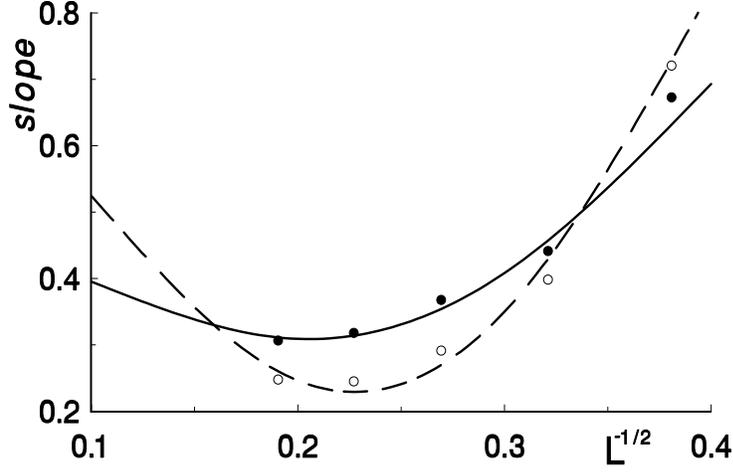,width=11cm,height=8.5cm}}
\vspace{-3ex}
\caption{\small Slope of the $y_{eff}''$ vs $L^{-1/2}$ plot in
Fig.~\ref{zeroth} (including also smaller sizes $L$). The empty
circles correspond to $\phi = \exp(-\beta)$, whereas the solid
circles to $\phi = \sinh( 2 \beta)$. The corresponding least--squares
fits $1.1840 - 8.507 L^{-1/2} +19.09 L^{-1}$ and
$0.6669 - 3.645 L^{-1/2} + 9.275 L^{-1}$ are shown by long--dashed
line and solid line, respectively.}
\label{slope}
\end{figure}

\begin{figure}
\centerline{\psfig{figure=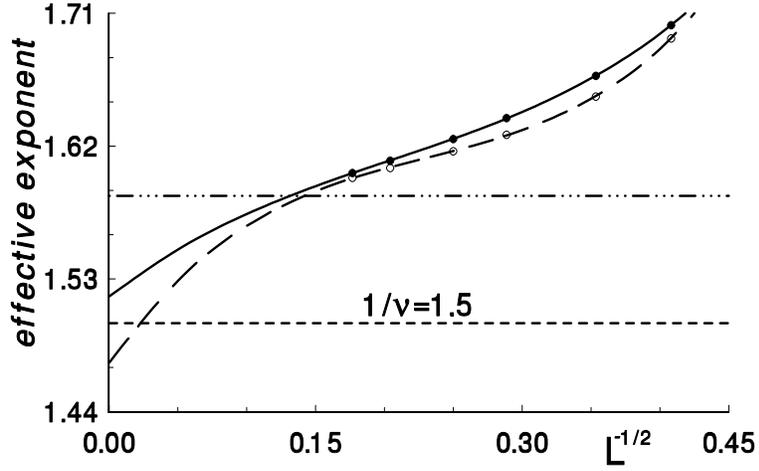,width=11cm,height=8.5cm}}
\vspace{-3ex}
\caption{\small Effective critical exponent $y_{eff}''(L)$
for the imaginary part of the complex partition--function--zeroth
%of 3D Ising model
as a function of $L^{-1/2}$, where $L$ is the linear size of the system.
The empty circles correspond to $\phi = \exp(-\beta)$, whereas the solid
circles to $\phi = \sinh( 2 \beta)$. The corresponding least--squares
fits $y_{eff}''(L)= 1.4731 + 1.3345 L^{-1/2} - 4.7657 L^{-1}
+ 6.8962 L^{-3/2}$ and
$y_{eff}''(L)= 1.5180 + 0.7301 L^{-1/2} - 2.0397 L^{-1}
+ 3.3181 L^{-3/2}$ are shown by long--dashed line
and solid line, respectively. 
Our asymptotic value $y''=1/\nu=1.5$ is indicated by horizontal
dashed line, whereas that of the RG theory ($1.5863$)
-- by dot--dot--dashed line.}
\label{zeroref}
\end{figure}

The data points of $y_{eff}'$ look (and are expected to be) less accurate
than those of $y_{eff}''$, since $Re \left( u_1^0-u_c \right)$
has a very small value.
The $y_{eff}''$ data do not look scattered, therefore they allow a refined
analysis with account for nonlinear corrections. To obtain stable results,
we have included the data for smaller lattice sizes $L=3$ and $L=4$ given
in~\cite{ABV}. In principle, we can use rather arbitrary analytical function
$\phi (\beta)$ to evaluate the effective critical exponent
$$y_{eff}''(L) = \ln \left[ Im \, \phi \left(\beta_1^0(L/2) \right)
/ Im \, \phi \left( \beta_1^0(L) \right) \right] / \ln 2$$
and estimate its asymptotic value $y''$. For an optimal choice, however,
$y_{eff}''(L)$ vs $L^{-\omega}$ plot should be as far as possible linear
to minimize the extrapolation error. In this aspect, our choice
$\phi = \exp(-\beta)$ is preferable to $\phi = \exp(-4 \beta)$ used
in~\cite{ABV}. We have tested also another possibility, i.~e.
$\phi = \sinh( 2 \beta)$, providing almost optimal results in
the case of 2D Ising model. In Fig.~\ref{slope} we have shown the
slope of $y_{eff}''$ vs $L^{-1/2}$ curve, calculated from the MC data
of~\cite{ADH,ABV}, for $\phi = \exp(- \beta)$ (empty circles) and
\linebreak $\phi = \sinh( 2 \beta)$ (solid circles).
It is evident that in both
cases the slope cannot be reasonably approximated by a linear
function of $L^{-1/2}$, but can be quite well described by
a parabola. The latter means that $y_{eff}''(L)$ can be satisfactory
well approximated by a third--order (but not by a second--order)
polinomial in $L^{-1/2}$. The
corresponding four parameter least--squares fits are shown in
Fig.~\ref{zeroref}.
They yield $y'' \simeq 1.473$ in the case of
$\phi = \exp(-\beta)$ (long--dashed line) and
$y'' \simeq 1.518$ at $\phi = \sinh(2 \beta)$ (solid line).
It is evident from Fig.~\ref{zeroref} that
in the latter case we have slightly better linearity of the fit,
therefore $1/\nu \simeq 1.518$ is our best estimate of 
the critical exponent $1/\nu$ from the actual MC data.
Thus, while the row estimation provided the value
$y''=1/\nu \simeq 1.552$ which is closer to the RG 
prediction $1/\nu \simeq 1.5863$ (horizontal dot--dot--dashed line),
the refined analysis reveals remarkably better agreement with our
(exact) value $1/\nu=1.5$ (horizontal dashed line).

\section{Conclusions}

In summary, we conclude the following.
\begin{enumerate}
\item Corrections to scaling for different physical quantities
near and at criticality have been discussed in framework of our recently
developed theory~\cite{my3} (Sec.~\ref{sec:crex}).
\item The critical exponent $\gamma$ for 3D Heisenberg model
has been estimated by
fitting the original susceptibility (MC) data of~\cite{Janke} to an
ansatz of finite--size--scaling theory which includes the leading
confluent correction--to--scaling term
(Sec.~\ref{sec:gam}). The obtained
estimates ($\gamma=1.345 \pm 0.023$ and $\gamma = 1.354 \pm 0.020$)
agree within error bars with
our theoretical value $\gamma=19/14 \simeq 1.35714$ and disagree with the
conventional RG value $\gamma \simeq 1.3895$. Taking into account
also the leading analytical correction, a selfconsistent 
estimation always yields the central value of $\gamma$ in the range of
$1.345 \le \gamma \le 1.369$, i.~e., reasonably close to our (exact) value $19/14$,
if the ratio of amplitudes $r$ for analytical and confluent corrections
is varied from $0$ to $\infty$. 
\item  Based on MC data for susceptibility in
3D Heisenberg model, a very accurate estimation of the critical coupling
has been made at given values of critical exponents
(Sec.~\ref{sec:coupl}), taking into account both confluent and analytical
corrections to scaling. These estimates, combined with fits
in vicinity of the critical point (Sec.~\ref{sec:test}),
allowed us to test the consistency between theoretical values of critical
exponents and actual MC data. As a result, we have found that
our values ($\eta=1/10$, $\gamma=19/14$, $\omega=3/5$)
are consistent, whereas those of the RG theory
($\eta \simeq 0.0355$, $\gamma \simeq 1.3895$, $\omega \simeq 0.782$)
are rather inconsistent with the MC data.
\item Recent Monte Carlo data for complex zeroth of the partition function
in 3D Ising model have been discussed (Sec.~\ref{sec:zeroth}).
The actual MC data suggest that, like in 2D Ising model, the critical
behavior of the real part differs from that of the imaginary part.
It can be explained reasonably by our exponents $\nu=2/3$ and $\omega=1/2$,
but not by those ($\nu \simeq 0.6304$ and $\omega \simeq 0.799$) of
the conventional RG theory. Our best estimate of the critical
exponent $\nu$ from the MC data, i.~e. $1/\nu \simeq 1.518$ or
$\nu \simeq 0.659$, is in a good agreement with the theoretical
(exact) value $2/3$.
\end{enumerate}

\end{document}